\documentclass[11pt]{article}

\usepackage[final]{acl}

\usepackage{times}
\usepackage{latexsym}

\usepackage[T1]{fontenc}

\usepackage[utf8]{inputenc}

\usepackage{microtype}

\usepackage{inconsolata}

\usepackage{graphicx}

\usepackage{booktabs}
\usepackage{wrapfig}
\usepackage{algorithm}
\usepackage{algorithmicx}
\usepackage{subfigure}
\usepackage{amsmath}
\usepackage{amssymb}
\usepackage{mathtools}
\usepackage{amsthm}
\usepackage{algpseudocode}
\usepackage{adjustbox}
\usepackage{makecell}
\usepackage{multirow}
\usepackage[capitalize,noabbrev]{cleveref}
\usepackage{balance}

\usepackage{fancyhdr}

\fancypagestyle{emnlpfirst}{
    \fancyhf{}
    \fancyfoot[L]{\footnotesize Accepted to EMNLP 2026.}

}

\title{Learning When to Think: Utility-Guided Adaptive\\Reasoning for Efficient Text Retrieval}

\author{
  Yingtao Luo\textsuperscript{1} \quad
  Kangyi Zhao\textsuperscript{2} \quad
  Shiyi Du\textsuperscript{1} \quad
  Yubo Li\textsuperscript{1} \\
  \textsuperscript{1}Carnegie Mellon University \quad
  \textsuperscript{2}University of Pittsburgh \\
  \texttt{\{yingtaol, shiyid, yubol\}@andrew.cmu.edu} \\
  \texttt{kaz78@pitt.edu}
}

\begin{document}
\maketitle
\thispagestyle{emnlpfirst}
\begin{abstract}
Reasoning-conditioned embeddings can improve retrieval by progressively refining semantic constraints. However, additional reasoning may introduce over-generalized or semantically noisy abstractions that weaken retrieval specificity. Existing methods largely supervise reasoning through final retrieval outcomes, providing limited guidance over the refinement process itself. We propose AURA, a framework that directly supervises intermediate reasoning states according to how well they separate relevant documents from hard distractors. AURA further trains a lightweight utility estimator over prefix-conditioned embeddings to adapt reasoning depth at inference time. On the reasoning-intensive BRIGHT benchmark, AURA-4B achieves 34.7 average nDCG@10 while using 2.8 reasoning steps and 96 generated tokens per query. Under matched deployment conditions, adaptive stopping reduces latency from 4.6 to 0.76 seconds relative to full-trajectory reasoning. Results on MS MARCO further suggest that this specialization need not substantially degrade conventional in-domain retrieval.
\end{abstract}

\section{Introduction}

Recent dense retrieval systems increasingly incorporate reasoning through query rewriting, decomposition, reranking, and reasoning-conditioned embeddings~\citep{qin2025tongsearchqrreinforcedqueryreasoning,jiang2025deepretrievalhackingrealsearch,asai2024self,li2025r3,muennighoff2024generative,tang2025large,lan2025ume}, motivated by the observation that semantic refinement can improve retrieval for complex queries.

\begin{figure}
    \centering
    \includegraphics[width=1\linewidth]{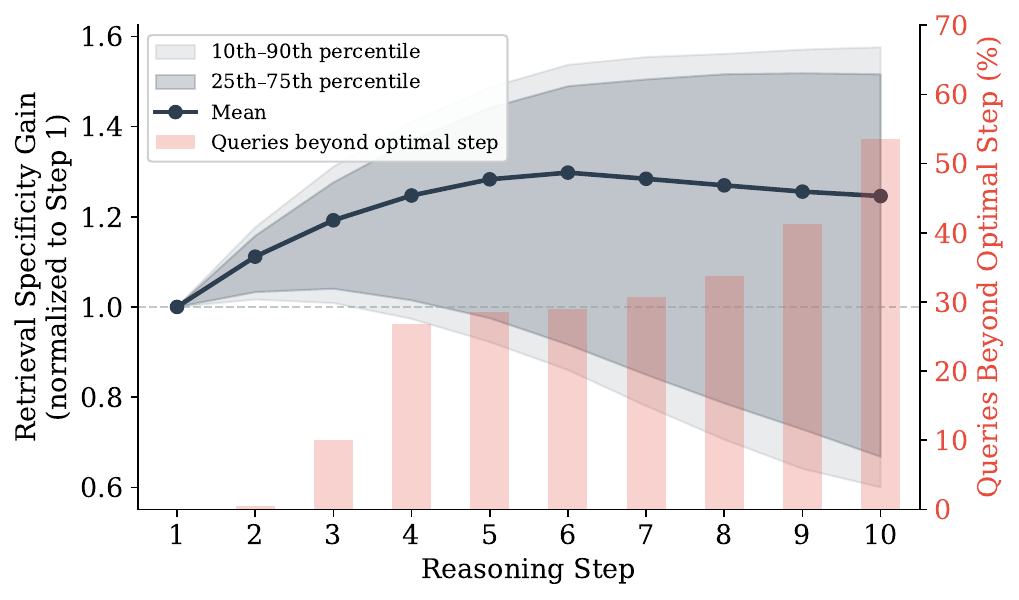}
    \caption{Retrieval specificity across reasoning steps on BRIGHT. Lines and shaded regions show the mean and query-level variability normalized to Step~1. Bars indicate the fraction of queries that have already passed their query-specific optimal step.}
    \label{fig:utility_dynamics}
\end{figure}

However, existing reasoning-based retrieval systems generally treat additional reasoning as beneficial, applying it without distinguishing whether further refinement helps or hurts retrieval quality. In practice, retrieval-oriented reasoning often exhibits diminishing returns. While some refinements sharpen \emph{retrieval specificity}---the separation between relevant documents and hard distractors---by identifying useful constraints or disambiguating entities, others introduce semantic drift or overly abstract reformulations that weaken retrieval discrimination~\citep{fan2026tfrank,kotte2026not}. As a result, longer reasoning trajectories do not always improve retrieval quality, and can even harm retrieval discrimination, increase inference latency, and introduce unnecessary computational cost.

To understand how retrieval utility evolves during reasoning, we analyze retrieval specificity across intermediate reasoning steps on BRIGHT. As shown in Figure~\ref{fig:utility_dynamics}, retrieval specificity improves on average during early reasoning steps, suggesting that additional refinement can help construct more discriminative retrieval representations. However, this improvement is highly heterogeneous: variability increases rapidly with reasoning depth, indicating that different queries benefit from substantially different amounts of reasoning. The fraction of queries that have already passed their optimal reasoning step increases and exceeds 50\% by Step~10. This evidence supports the notion that additional reasoning can be unnecessary and may even reduce the retrieval quality of many queries. Effective retrieval requires adapting the depth of the reasoning to the query-specific need.

\begin{figure}[t]
    \centering
    \includegraphics[width=1\linewidth]{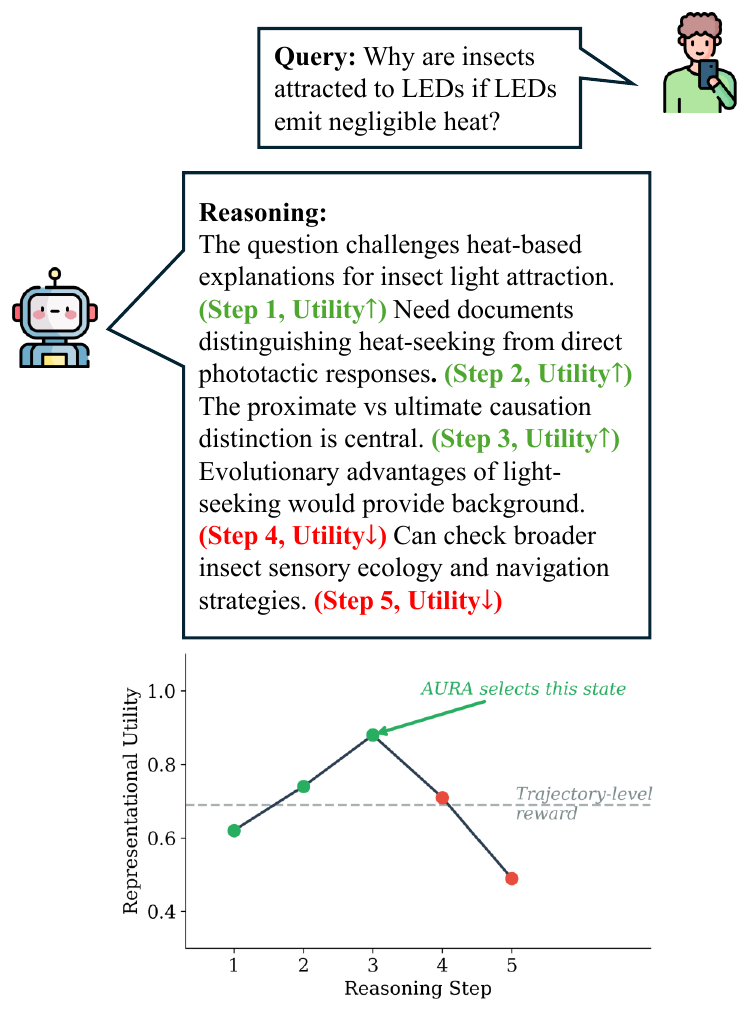}
    \caption{Motivation behind AURA. Existing methods supervise only the final retrieval outcome, while AURA identifies the peak-utility reasoning state before over-reasoning weakens retrieval specificity.}
    \label{fig:concept}
\end{figure}

\textbf{A key limitation of existing methods is that reasoning is typically supervised only through final retrieval performance}. While effective for evaluating end-to-end success, these coarse signals provide little guidance over the reasoning process. More importantly, current retrieval systems cannot predict whether additional reasoning will improve retrieval before observing retrieval outcomes, and thus can only allocate computation reactively: they expand reasoning trajectories regardless of whether further refinement is likely to help.

In this work, we formulate retrieval-oriented reasoning as a utility-guided computation allocation problem. Rather than treating reasoning as an unconditional token expansion, we evaluate whether each intermediate refinement improves the model's ability to distinguish relevant documents from challenging distractors (Figure~\ref{fig:concept}). Under this perspective, useful reasoning is determined not by trajectory length or textual sophistication, but by its measurable effect on retrieval discrimination.

Building on this perspective, we propose \textbf{AURA}\footnote{Project repository: \url{https://github.com/yingtaoluo/Adaptive-Reasoning-Retrieval}.} (\textbf{A}daptive \textbf{U}tility-aware \textbf{R}easoning Retriev\textbf{A}l), a framework for learning retrieval-oriented reasoning through utility supervision over intermediate reasoning states. During training, AURA uses relevance labels and retrieval candidates to estimate whether each refinement step improves retrieval discrimination, encouraging reasoning trajectories that sharpen retrieval specificity while avoiding semantically noisy refinements.

Since retrieval specificity is unavailable at inference time, AURA additionally trains a lightweight utility estimator to predict the retrieval utility of each current reasoning state. Monitoring this estimate online allows AURA to stop when state utility ceases to improve, without exhaustively expanding the reasoning trajectory.

We evaluate AURA on the 12 reasoning-intensive retrieval domains in BRIGHT~\citep{hongjinbright} and on MS MARCO~\citep{bajaj2016ms} as a conventional in-domain retrieval setting. AURA outperforms the compared reasoning-conditioned retrieval methods on BRIGHT. On MS MARCO, it remains comparable to a strong reasoning-trained retriever, suggesting no substantial in-domain degradation when extensive reasoning is unnecessary. Controlled analyses further show that the gains cannot be explained by denser intermediate contrastive supervision or trajectory filtering alone, and that retrieval specificity is strongly aligned with corpus-level nDCG@10.

Our contributions are summarized as follows:
\begin{itemize}
\item We formulate reasoning-conditioned retrieval as an adaptive refinement process and show that reasoning utility is highly heterogeneous.
\item We propose \textbf{AURA}, a framework that supervises intermediate reasoning states according to their retrieval specificity, and trains a lightweight utility estimator to dynamically allocate reasoning during inference.
\item We demonstrate improvements over the compared methods across 12 reasoning-intensive domains, together with a substantially better performance--cost trade-off under matched inference conditions.
\end{itemize}

\section{Related Work}
\label{related}

\subsection{Dense Retrieval and Embedding Models}
Modern dense retrieval commonly relies on bi-encoder architectures that map queries and documents into a shared embedding space, where retrieval is performed through vector similarity~\citep{karpukhin2020dense}. Early embedding approaches such as Sentence-BERT~\citep{reimers2019sentence} demonstrated the effectiveness of contrastive representation learning for semantic retrieval, followed by stronger instruction-tuned and large embedding models including INSTRUCTOR~\citep{su2023one}, E5~\citep{wang2022text}, and recent LLM-based embedding systems~\citep{muennighoff2024generative, liu2024codexembed, wang2024improving}. Recent studies further explore using LLM hidden states directly as embeddings~\citep{ma2024fine,lee2025nvembed} or leveraging in-context learning to improve representation quality~\citep{ICLR2025_5eba8556}. 
Beyond supervised retrieval data, dense retrievers have also been trained using unsupervised contrastive objectives and synthetic supervision~\citep{lee2019latent, izacard2021unsupervised, chen2024bge, ji2025learning, yoon2025square}. Despite strong performance, most embedding models remain fundamentally single-pass encoders that construct representations without explicit semantic refinement or intermediate reasoning.

\subsection{Reasoning for Retrieval} 

Recent work increasingly incorporates reasoning into retrieval systems through query rewriting, decomposition, iterative retrieval, and reasoning-augmented reranking~\citep{qin2025tongsearchqrreinforcedqueryreasoning,jiang2025deepretrievalhackingrealsearch,asai2024self,li2025r3,long2025divermultistageapproachreasoningintensive}. For example, ReasonIR~\citep{shao2025reasonirtrainingretrieversreasoning} synthesizes reasoning trajectories and hard negatives to train retrieval models on reasoning-intensive benchmarks~\citep{hongjinbright}. More recently, reasoning has been incorporated directly into embedding formation: multimodal methods such as RITE~\citep{liu2025exploring}, LREM~\citep{tang2025large}, and UME-R1~\citep{lan2025ume} generate reasoning before constructing retrieval representations, with further gains from inference-time scaling through sampling multiple reasoning trajectories~\citep{muennighoff2024generative}.

However, reasoning is not uniformly beneficial for retrieval. Query expansion has long been known to suffer from query drift~\citep{mitra1998improving,carpineto2012survey}, and recent work shows that LLM-based reasoning can hurt dense retrieval due to semantic drift~\citep{fan2026tfrank,kotte2026not}. Similar observations have emerged for chain-of-thought prompting more broadly~\citep{liu2025mind,hassid2025don,suistop}. Despite these findings, existing methods either supervise reasoning only through final retrieval outcomes or select among complete trajectories via best-of-$k$, providing no direct supervision over intermediate refinement states nor adaptive reasoning allocation at inference time. AURA addresses both gaps.

\begin{figure*}[t]
\centering
\includegraphics[width=0.99\linewidth]{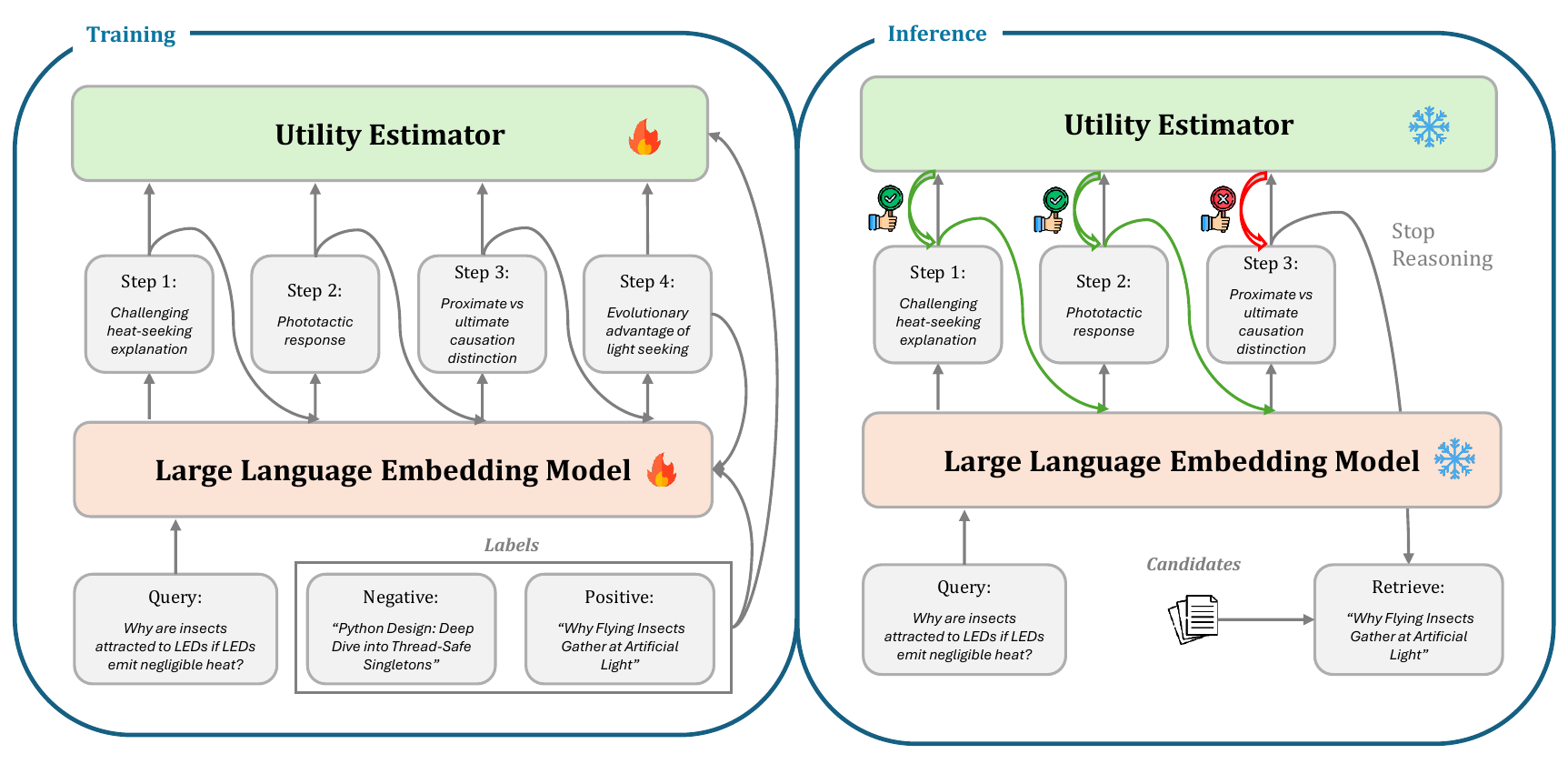}
\caption{Overview of AURA. \textbf{Training:} A warm-up stage first establishes a retrieval-oriented embedding space. Specificity labels are then constructed from the warmed-up model, the utility estimator is trained on frozen embeddings, and utility-aware weighting is enabled for subsequent joint optimization. \textbf{Inference:} The utility estimator evaluates each state online and terminates reasoning according to the predicted utility, returning the selected embedding for retrieval.}
\label{fig:diagram}
\end{figure*}

\section{Methodology}

\subsection{Unified Reasoning Embedding}
To ground reasoning-conditioned retrieval within a unified framework, AURA utilizes an autoregressive language model as its unified backbone $f_\theta$. 
To enable a single language model to simultaneously optimize autoregressive token generation and extract high-quality dense retrieval representations, we propose a structured sequence format. Given a query $x$, the format is as follows:
\begin{center}
\label{box:template}
\footnotesize
\framebox[\linewidth][l]{
\begin{minipage}{0.96\linewidth}
\textbf{Unified Sequence Format:} \\
\texttt{[User]:} Input: $x$. Generate step-by-step reasoning. \\
\texttt{[Assistant]:} \texttt{<think>} $r_1, r_2, \dots, r_T$ \texttt{</think>}
\end{minipage}
}
\end{center}
\noindent where $r = (r_1, \dots, r_T)$ represents the generated reasoning trajectory parsed at sentence boundaries.

\paragraph{Embedding extraction.} At each sentence boundary $t$, we construct a prefix-conditioned embedding by passing the query and reasoning prefix through the backbone and extracting the final-layer hidden state at the last token position of $r_t$:
\begin{equation}
z_t = \mathrm{proj}(\mathrm{LastHidden}_\theta(x, r_{\le t})),
\end{equation}
where $\mathrm{LastHidden}_\theta(\cdot)$ returns the final layer's hidden state at the last token of the input, and $\mathrm{proj}(\cdot)$ is a linear projection to the embedding space. 

\paragraph{Retrieval Specificity Scoring.} To supervise intermediate reasoning states, AURA defines a retrieval specificity score that measures how well a given reasoning state separates relevant documents from distractors.
For a training query $x$, let $d^+$ denote a relevant document and let $\mathcal{D}^-(x)$ denote the set of in-batch negatives. Given $z_t$, we define its retrieval specificity score as
\begin{equation}
S(z_t) = \mathrm{sim}(z_t, z_{d^+}) - \frac{1}{k}\sum_{i=1}^{k}\mathrm{sim}(z_t, z_{d_i^-}),
\end{equation}
where $d_1^-, \dots, d_k^-$ are the top-$k$ highest-similarity distractors from $\mathcal{D}^-(x)$ according to $\mathrm{sim}(\cdot, \cdot)$ as cosine similarity. $S(z_t)$ evaluates the retrieval specificity at each intermediate reasoning state. The marginal effect of a reasoning step can then be measured as $\Delta S_t = S(z_t) - S(z_{t-1})$.

\paragraph{Training objectives.} The model is trained with a multi-task loss combining contrastive retrieval and language modeling objectives:
\begin{equation}
\mathcal{L}_{\mathrm{total}} = \mathcal{L}_{\mathrm{InfoNCE}} + \mathcal{L}_{\mathrm{CE}}
\end{equation}

The autoregressive objective is computed over all reasoning tokens, preserving the model's ability to generate coherent reasoning:
\begin{equation}
\mathcal{L}_{\mathrm{CE}} = -\sum_{t=1}^{T} \log P_\theta(r_t \mid x, r_{<t})
\end{equation}

The contrastive objective optimizes the prefix-conditioned embeddings to separate relevant documents from distractors. Unlike existing methods that apply InfoNCE only at the final state $z_T$, AURA optimizes at all intermediate states $z_1, \dots, z_T$. For each intermediate embedding $z_t$, we compute the InfoNCE loss against the relevant document embedding $z_{d^+}$ and in-batch negatives:
\begin{equation}
\mathcal{L}_{\mathrm{InfoNCE}} = -\frac{1}{T}\sum_{t=1}^{T} w_t \log \frac{\exp(z_t \cdot z_{d^+} / \tau)}{\sum_{d \in \mathcal{D}} \exp(z_t \cdot z_d / \tau)}
\end{equation}
where $\mathcal{D}=
\{d^+\}\cup\{d^-_j\}$ contains the relevant document and all in-batch negatives, $\tau$ is the contrastive temperature (default set to 0.02), and the weight is set as 
\begin{equation}
w_t=
\frac{
\exp(
\max(0,\mathrm{stopgrad}(\Delta S_t))
)
}{
\sum_{t'}
\exp(
\max(0,\mathrm{stopgrad}(\Delta S_{t'}))
)
},
\end{equation}
which emphasizes the learning of reasoning refinements that produce larger gains in retrieval specificity and reduces supervision on saturated or retrieval-harmful reasoning states. Here, we apply stop-gradient so that utility scores are used only to reweight contrastive updates, preventing the model from artificially increasing utility values instead of improving retrieval representations.

\paragraph{Gradient isolation.}
Since both objectives share the same backbone model and operate on overlapping reasoning positions, naive joint optimization may cause contrastive gradients to interfere with autoregressive generation. To mitigate this effect, during the contrastive optimization, we apply stop-gradient to all positions preceding the embedding extraction point, such that gradients from $\mathcal{L}_{\mathrm{InfoNCE}}$ do not propagate through $(x,r_{<t})$.

Under this design, $\mathcal{L}_{\mathrm{CE}}$ retains full gradient flow across the reasoning sequence, while $\mathcal{L}_{\mathrm{InfoNCE}}$ only updates representations at the current reasoning step without backpropagating through earlier reasoning prefixes. This separation reduces optimization interference between generation and retrieval objectives, allowing the model to learn discriminative retrieval representations while preserving autoregressive reasoning quality.

\subsection{Inference-Time Utility Estimation}
Since retrieval specificity $S(z_t)$ depends on relevance labels and distractors, it is unavailable at inference time.
To enable adaptive reasoning, AURA learns a utility estimator that predicts the retrieval specificity of the current intermediate state.
For each state $z_t$, we define its target utility as
\begin{equation}
y_t
=
S(z_t),
\end{equation}
which represents the observed state-level retrieval utility.
The estimator
\begin{equation}
V_\phi(z_t)
\end{equation}
is trained to preserve the ordering of state utility using a pairwise ranking loss:
\begin{equation}
\mathcal L_{\mathrm{rank}}
=
-
\log
\sigma\!\left(
V_\phi(z_i)-V_\phi(z_j)
\right),
\qquad
y_i>y_j.
\end{equation}
Since adaptive stopping and state selection depend only on the relative ordering of reasoning states within a query trajectory, this objective targets ordinal agreement rather than absolute score calibration and is less sensitive to query- and negative-set-specific shifts in the scale of retrieval specificity.
At inference time, reasoning terminates when the predicted state utility has not improved for $p$ consecutive steps, and the generated state with the highest predicted utility is used for retrieval.
The utility estimator is a lightweight MLP over prefix-conditioned embeddings.

\section{Experiments}
\label{exp}

\begin{table*}[t]
\centering
\small
\setlength{\tabcolsep}{6pt}
\caption{nDCG@10 on BRIGHT under three inference settings.}
\label{tab:main_results}
\begin{adjustbox}{width=0.99\textwidth}
\begin{tabular}{l|c|ccccccc|cc|ccc}
\toprule
Method & Avg. 
& \multicolumn{7}{c|}{StackExchange} 
& \multicolumn{2}{c|}{Coding} 
& \multicolumn{3}{c}{Theorem-based} \\
\cline{3-14}
& 
& Bio. & Earth. & Econ. & Psy. & Rob. & Stack. & Sus. 
& Leet. & Pony 
& AoPS & TheoQ. & TheoT. \\
\midrule
\multicolumn{14}{l}{\textit{\textbf{Setting I: Single-pass retrieval with original query}}} \\
\midrule
BM25~\citep{robertson2009probabilistic} & 14.5 & 18.9 & 27.2 & 14.9 & 12.5 & 13.6 & 18.4 & 15.0 & 24.4 & 7.9 & 6.2 & 10.4 & 4.9 \\
SBERT~\citep{reimers2019sentence} & 14.9 & 15.1 & 20.4 & 16.6 & 22.7 & 8.2 & 11.0 & 15.3 & 26.4 & 7.0 & 5.3 & 20.0 & 10.8 \\
Qwen3-4B~\citep{zhang2025qwen3embeddingadvancingtext} & 5.6 & 3.5 & 8.0 & 2.3 & 2.0 & 1.6 & 1.0 & 4.4 & 2.1 & 0.1 & 4.9 & 18.0 & 19.2 \\
OpenAI text-embedding-3-large & 17.9 & 23.3 & 26.7 & 19.5 & 27.6 & 12.8 & 14.3 & 20.5 & 23.6 & 2.4 & 8.5 & 23.5 & 11.7 \\
ReasonIR-8B~\citep{shao2025reasonirtrainingretrieversreasoning} & 24.4 & 26.2 & 31.4 & \textbf{23.3} & 30.0 & \underline{18.0} & \textbf{23.9} & 20.5 & 35.0 & \underline{10.5} & \textbf{14.7} & \underline{31.9} & 27.2 \\
RaDeR-7B~\citep{das2025rader} & \underline{25.5} & \underline{34.6} & \underline{38.9} & \underline{22.1} & \underline{33.0} & 14.8 & \underline{22.5} & \underline{23.7} & \underline{37.3} & 5.0 & 10.2 & 28.4 & \underline{35.1} \\
DIVER-4B~\citep{long2025divermultistageapproachreasoningintensive} & \textbf{28.9} & \textbf{41.8} & \textbf{43.7} & 21.7 & \textbf{35.3} & \textbf{21.0} & 21.2 & \textbf{25.1} & \textbf{37.6} & \textbf{13.2} & 10.7 & \textbf{38.4} & \textbf{37.3} \\
\midrule
\multicolumn{14}{l}{\textit{\textbf{Setting II: Retrieval with external GPT-generated query expansion}}} \\
\midrule
BM25 & 27.0 & \textbf{53.6} & \textbf{54.1} & 24.3 & 38.7 & 18.9 & \textbf{27.7} & \underline{26.3} & 19.3 & \underline{17.6} & 3.9 & 19.2 & 20.8 \\
SBERT & 17.8 & 18.5 & 26.3 & 17.5 & 27.2 & 8.8 & 11.8 & 17.5 & 24.3 & 10.3 & 5.0 & 22.3 & 23.5 \\
Qwen3-4B & 5.5 & 1.3 & 17.3 & 2.5 & 6.2 & 1.0 & 4.8 & 4.5 & 3.0 & 5.9 & 0.0 & 7.2 & 12.5 \\
OpenAI text-embedding-3-large & 23.3 & 35.2 & 40.1 & 25.1 & 38.0 & 13.6 & 18.2 & 24.2 & 24.5 & 6.5 & 7.7 & 23.8 & 22.9 \\
ReasonIR-8B & \underline{29.9} & 43.6 & 42.9 & \textbf{32.7} & \underline{38.8} & \underline{20.9} & 25.8 & \textbf{27.5} & 31.5 & \textbf{19.6} & 7.4 & 33.1 & 35.7 \\
RaDeR-7B & 29.2 & 36.1 & 42.9 & 25.2 & 37.9 & 16.6 & 27.4 & 25.0 & \textbf{34.8} & 11.9 & \textbf{12.0} & \underline{37.7} & \textbf{43.4} \\
DIVER-Retriever-4B & \textbf{32.1} & \underline{51.9} & \underline{53.5} & \underline{29.5} & \textbf{41.2} & \textbf{21.4} & \underline{27.5} & 26.1 & \underline{33.5} & 11.7 & \underline{9.5} & \textbf{39.3} & \underline{39.7} \\
\midrule
\multicolumn{14}{l}{\textit{\textbf{Setting III: Retrieval with adaptive internal reasoning (w/o external LLM)}}} \\
\midrule
AURA-4B (base) & \textbf{34.7} & \textbf{47.5} & \textbf{48.6} & \textbf{28.3} & \textbf{43.5} & \textbf{23.8} & \textbf{31.2} & \textbf{34.6} & \textbf{40.9} & \textbf{14.2} & \textbf{17.0} & \textbf{43.2} & \textbf{43.3} \\
\bottomrule
\end{tabular}
\end{adjustbox}
\end{table*}

\subsection{Training Data Construction}

We build on the publicly released training data with 101K query-positive-negative samples from ReasonIR~\citep{shao2025reasonirtrainingretrieversreasoning}, which provides reasoning-intensive queries with paired positive documents and hard negatives synthesized from MS MARCO\footnote{\url{https://huggingface.co/datasets/microsoft/ms_marco}} seed documents. 
To obtain reasoning trajectories for utility supervision, we sample $G{=}8$ independent trajectories per query from DeepSeek-R1~\citep{guo2025deepseek}, each truncated to a maximum of $l{=}2{,}048$ tokens. Each trajectory is segmented at sentence boundaries into intermediate reasoning states. After the retrieval warm-up described below, we compute retrieval specificity $S(z_t)$ for every intermediate state using the labeled positive document and hard negatives provided by ReasonIR. Trajectories where no intermediate state improves over the initial query embedding are removed during label construction, retaining 67.5\% of sampled trajectories (approximately 5.4 per query on average), for approximately 545K reasoning samples.

\subsection{Training Procedure}
Training separates label construction from utility-aware optimization. First, we warm up the unified model for one epoch with the multi-task objective but without utility-guided weighting. We freeze this warmed-up checkpoint to construct retrieval-specificity and state-utility labels. The utility estimator $V_\phi$, a 3-layer 10M-parameter MLP with ReLU activations, is trained on these frozen embeddings with a per-GPU batch size of 256 and a learning rate of $5 \times 10^{-4}$ for up to ten epochs, with early stopping based on loss convergence. We then enable utility-aware weighting and continue joint optimization of the unified model. AdamW is used throughout. This schedule avoids constructing labels from the backbone's initial, non-retrieval-adapted representations.

\subsection{Implementation Details}
Experiments are conducted on 8 NVIDIA H100 GPUs, consuming approximately 100 GPU hours to converge. We adopt Qwen3-4B-Instruct as the backbone model and use a per-GPU batch size of 128 and a learning rate of $1 \times 10^{-5}$. The embedding dimension matches the backbone hidden size. We primarily report nDCG@10, together with generated tokens, average reasoning steps, and matched-setting wall-clock latency.

\subsection{Baselines}
We organize baselines into three categories according to whether
inference-time reasoning is used and how reasoning is integrated
with retrieval.

Setting I evaluates standard retrieval systems without explicit
inference-time reasoning, where representations are constructed
directly from the original query.
Setting II evaluates retrieval augmented with external query
rewriting, which follows the decoupled paradigm of recent reasoning-intensive retrieval works, where reasoning is generated by an external LLM at inference time.

To provide strong baselines, we intentionally adopt GPT-4 for query expansion rather than external reasoners with comparable size to our Qwen3-4B backbone. We also evaluate Qwen3-4B-Instruct, AURA's backbone, as an external query rewriter using a prompt comparable to the GPT-4 baseline. As shown in Table~\ref{tab:self_rewrite}, it does not improve any evaluated retriever and reduces average BRIGHT nDCG@10 by 0.1--2.7 points. We therefore use GPT-4 rewriting to construct a stronger external-reasoning baseline.

\begin{table}[t]
\centering
\small
\caption{Effect of Qwen3-4B external query rewriting on BRIGHT (average nDCG@10).}
\label{tab:self_rewrite}
\begin{tabular}{lcc}
\toprule
\textbf{Retriever} & \textbf{Original} & \textbf{+ Rewrite} \\
\midrule
BM25 & 14.5 & 12.9 \\
SBERT & 14.9 & 12.6 \\
Qwen3-4B & 5.6 & 5.5 \\
OpenAI-3-large & 17.9 & 15.2 \\
\bottomrule
\end{tabular}
\end{table}

Setting III evaluates endogenous reasoning, the new paradigm of AURA, where reasoning and
embedding formation are performed within the same model. Our
method further introduces utility estimation to adaptively
allocate inference-time reasoning and avoid unnecessary
refinement.

\subsection{Main Results}
Table~\ref{tab:main_results} reports retrieval performance of different models measured by nDCG@10 on BRIGHT benchmark~\citep{hongjinbright}.

First, Setting II consistently outperforms Setting I across
retrievers. This result is aligned with recent reasoning-intensive
retrieval literature and suggests that augmenting retrieval with
external query rewriting can substantially improve retrieval
quality. Notably, Setting II baselines use a separate external
reasoning model (GPT-4) for query expansion while keeping the
retriever unchanged. Across methods, this external reasoning strategy
appears to provide the largest improvements for lexical
retrievers such as BM25.

AURA achieves the best overall performance among the compared systems with an average
nDCG@10 of 34.7. Despite not relying on any external reasoning
model, AURA outperforms all Setting II baselines that are powered by
GPT-4 query expansion. The gains are especially pronounced in
the theorem-based subsets, where AURA shows the largest margin
over existing approaches. The strongest external reasoning
baseline, DIVER-Retriever-4B, achieves 32.1 with GPT-4 expanded
queries, while AURA improves this to 34.7 (+2.6) using reasoning
generated entirely within the retrieval model itself.

Compared with Setting I, the advantage of adaptive reasoning is
even larger. The strongest baseline without inference-time
reasoning achieves 28.9, whereas AURA reaches 34.7 (+5.8
absolute, +20.1\% relative). 

\subsection{Alignment with Corpus-Level Retrieval}
\label{sec:corpus_alignment}
Retrieval specificity is computed from labeled positives and hard negatives during training, whereas the final evaluation ranks the full corpus. To test whether the local signal reflects the listwise objective, we perform full-corpus BRIGHT retrieval for every available intermediate state in each naturally varying-length reasoning trajectory. For each query, we compute Spearman correlation across its states between retrieval specificity and corpus-level nDCG@10, and then average the query-level correlations. The resulting average correlation is 0.97. Moreover, the state selected by oracle specificity-based selection is also nDCG@10-optimal for 98.1\% of queries. This close agreement indicates that retrieval specificity tracks corpus-level ranking quality rather than only in-batch discrimination.

\subsection{Retrieval on Traditional IR Benchmarks}
Table~\ref{tab:msmarco} presents results on MS MARCO passage retrieval~\citep{bajaj2016ms}. We evaluate on the official development set, TREC-DL'19, and TREC-DL'20, all of which use held-out queries disjoint from the training split. These splits provide an in-domain check outside the reasoning-intensive BRIGHT setting. AURA-4B is comparable to RaDeR-7B across all three splits despite using a smaller backbone. This result suggests that utility-aware reasoning supervision does not substantially degrade conventional in-domain retrieval relative to a strong reasoning-trained retriever; broader general-purpose evaluation is left to future work.

\begin{table*}[t]
\centering
\small
\caption{Performance on MS MARCO passage retrieval (DEV and TREC DL19/DL20) as a standard retrieval setting rather than a reasoning-intensive one. Baseline results are from~\citep{das2025rader}.}
\label{tab:msmarco}
\begin{adjustbox}{width=0.7\linewidth}
\begin{tabular}{lcccc}
\toprule
\multirow{2}{*}{\textbf{Model}} & \multicolumn{2}{c}{\textbf{DEV}} & \textbf{DL19} & \textbf{DL20} \\
\cmidrule(lr){2-3} \cmidrule(lr){4-4} \cmidrule(lr){5-5}
 & MRR@10 & R@1k & nDCG@10 & nDCG@10 \\
\midrule
BM25~\citep{robertson2009probabilistic} & 18.4 & 85.3 & 50.6 & 48.0 \\
ANCE~\citep{xiong2020approximate} & 33.0 & 95.9 & 64.5 & 64.6 \\
CoCondenser~\citep{gao2022unsupervised} & 38.2 & 98.4 & 71.7 & 68.4 \\
OpenAI Ada2 & 34.4 & 98.6 & 70.4 & 67.6 \\
SimLM~\citep{wang2023simlm} & 41.1 & 98.7 & 71.4 & 69.7 \\
RepLLaMA~\citep{ma2024fine} & \textbf{41.2} & \textbf{99.4} & \textbf{74.3} & \textbf{72.1} \\
RaDeR-7B~\citep{das2025rader} & 34.4 & 98.1 & 71.2 & 70.7 \\
\midrule
AURA-4B (base) & 34.1 & 98.0 & 71.2 & 70.4 \\
\bottomrule
\end{tabular}
\end{adjustbox}
\end{table*}

\subsection{Ablation Study}

Table~\ref{tab:ablation} reports ablation results on BRIGHT. We examine five design dimensions of AURA, each targeting a specific question about the framework.

\paragraph{Q1: Is the specificity score robust to negative aggregation?}
The choice of negative aggregation in $S(z_t)$ has minimal impact: mean top-3, max, and mean all negatives yield similar performance (34.7, 34.6, 34.4). This indicates that AURA is robust to this design choice. Mean over all negatives may dilute the signal from hard negatives, while using only the hardest single negative may be susceptible to outlier noise; the default mean top-3 offers a reasonable balance.

\paragraph{Q2: Does utility-aware supervision improve over dense intermediate supervision?}
This is the central ablation: we ask whether the retrieval specificity signal provides meaningful supervision beyond simply applying contrastive training at more positions. We compare four different supervision strategies: (1)~\textit{final state only}, which applies InfoNCE only at $z_T$ as in prior work; (2)~\textit{random utility labels}, which trains the utility estimator with randomly assigned labels; (3)~\textit{monotonic utility labels}, which assumes reasoning quality increases monotonically with depth; and (4)~\textit{uniform-weight intermediate states}, which applies InfoNCE at all intermediate states with equal weight, removing utility-based differentiation.

\begin{table}[t]
\centering
\small
\caption{Ablation study evaluated by average nDCG@10 on BRIGHT benchmark. }
\label{tab:ablation}
\begin{adjustbox}{width=\columnwidth}
\begin{tabular}{lc}
\toprule
\textbf{Variant} & \textbf{nDCG@10} \\
\midrule
AURA (base) & 34.7 \\
\midrule
\multicolumn{2}{l}{\textit{Specificity scoring}} \\
\quad Mean top-3 negatives (default) & 34.7 \\
\quad $\max$ negative & 34.6 \\
\quad Mean all negatives & 34.4 \\
\midrule
\multicolumn{2}{l}{\textit{Training supervision}} \\
\quad All intermediate states (default) & 34.7 \\
\quad Final state only ($z_T$) & 31.2 \\
\quad Uniform-weight intermediate states & 33.6 \\
\quad Monotonic utility labels & 33.2 \\
\quad Random utility labels & 31.9 \\
\midrule
\multicolumn{2}{l}{\textit{Data construction and optimization}} \\
\quad Retain all trajectories & 34.6 \\
\quad Without gradient isolation & 29.8 \\
\midrule
\multicolumn{2}{l}{\textit{Inference-time utility estimation}} \\
\quad Utility estimator (default) & 34.7 \\
\quad LLM-confidence stopping (3.2 steps) & 32.8 \\
\quad Random selection & 32.3 \\
\quad Oracle specificity-based selection & 35.0 \\
\midrule
\multicolumn{2}{l}{\textit{Utility-aware stopping criterion}} \\
\quad $V_\phi$ with patience $p{=}1$ (default) & 34.7 \\
\quad Greedy peak stopping ($p{=}0$) & 34.6 \\
\quad $V_\phi$ with patience $p{=}3$ & 33.7 \\
\quad No early stopping (full trajectory) & 33.0 \\

\bottomrule
\end{tabular}
\end{adjustbox}
\end{table}

Results show a clear ordering: final-state-only (31.2) $<$ random (31.9) $<$ monotonic (33.2) $<$ uniform (33.6) $<$ AURA (34.7). Two findings are notable. First, uniform-weight supervision already improves substantially over final-state-only (+2.4), confirming that denser contrastive training is beneficial. However, AURA outperforms uniform weighting by a further 1.1 points, demonstrating that utility-aware differentiation among intermediate states provides supervision that cannot be achieved by simply increasing the number of contrastive training positions. Second, monotonic labels underperform uniform weighting ($-$0.4), indicating that the assumption of monotonically increasing reasoning quality is actively harmful---consistent with the over-reasoning phenomenon motivating this work. Random labels degrade close to final-state-only, confirming that the pairwise ranking loss requires accurate ordering signal to function effectively.

\paragraph{Q3: Are trajectory filtering and gradient isolation necessary?}
Retaining trajectories in which no intermediate state improves over the initial query changes nDCG@10 only marginally, from 34.7 to 34.6. Filtering therefore provides a small benefit by removing trajectories without a positive within-trajectory ranking signal, but does not explain AURA's overall improvement. In contrast, removing gradient isolation causes a substantial drop to 29.8. Retrieval gradients propagated through earlier prefixes therefore interfere with stable joint generation--retrieval optimization, supporting the gradient-isolation design.

\paragraph{Q4: How well does the utility estimator approximate oracle selection?}
At inference time, ground-truth retrieval specificity is unavailable, which motivates utility estimation. The utility estimator $V_\phi$ achieves 34.7, 0.3 points below the oracle upper bound (35.0), where the oracle selects the reasoning state with the highest true $S(z_t)$ using ground-truth relevance labels. Random state selection drops to 32.3. This narrow aggregate performance gap indicates that $V_\phi$ provides an effective approximation of retrieval utility without access to relevance labels.

\paragraph{Q5: Are simple adaptive stopping rules sufficient?}
We compare AURA with an LLM-confidence rule based on the mean token entropy of the current reasoning step, tuning both the threshold and stopping direction on development data. Its strongest variant obtains 32.8 nDCG@10 at 3.2 average steps, compared with AURA's 34.7 at 2.8 steps. Thus, generation confidence alone does not reliably identify the reasoning state that is most discriminative for retrieval.

\paragraph{Q6: When should reasoning stop?}
Early stopping consistently outperforms using the full reasoning trajectory (33.0), validating that over-reasoning degrades retrieval quality. Among stopping strategies, patience $p{=}1$ (34.7) performs best; greedy peak stopping ($p{=}0$) is comparable (34.6), while increasing patience to $p{=}3$ (33.7) allows too many low-utility steps through. The performance gap between $p{=}1$ and no early stopping (+1.7) confirms that adaptive termination is essential.

\subsection{Inference-Time Compute Efficiency}
 
Table~\ref{tab:compute_efficiency} compares inference-time reasoning cost. To isolate the effect of utility-guided stopping, we first compare AURA with and without early stopping under the same Qwen3-4B backbone, hardware, batch size, and software stack. Without early stopping, AURA generates 332 tokens (9.5 steps) per query, takes 4.6 seconds, and achieves 33.0 nDCG@10. Utility-guided stopping reduces generation to 96 tokens (2.8 steps) and latency to 0.76 seconds, corresponding to a 3.5$\times$ token reduction and approximately 6$\times$ speedup, while improving performance to 34.7 (+1.7). The utility estimator itself adds approximately 1 millisecond per query. For reference, the strongest external reasoning baseline (DIVER + GPT-4 expansion) generates 479 tokens per query using a separate reasoning model yet achieves lower performance (32.1). We do not report its wall-clock latency because external-model serving is not matched to our deployment.

\paragraph{Inference-time scaling through utility-guided reasoning search.}
Because AURA estimates retrieval utility at intermediate reasoning states, it naturally enables inference-time scaling by allocating additional reasoning computation only when further refinement is predicted to improve retrieval quality. We evaluate two search strategies under increasing inference budgets. 

(1) \textit{Best-of-$k$}: sample $k$ complete reasoning trajectories independently and select the trajectory with the highest estimated utility. 

(2) \textit{Greedy expansion}: generate multiple candidate refinements at each reasoning step and continue from the one with the highest predicted utility.

\begin{figure}[t]
\centering
\includegraphics[width=0.99\linewidth]{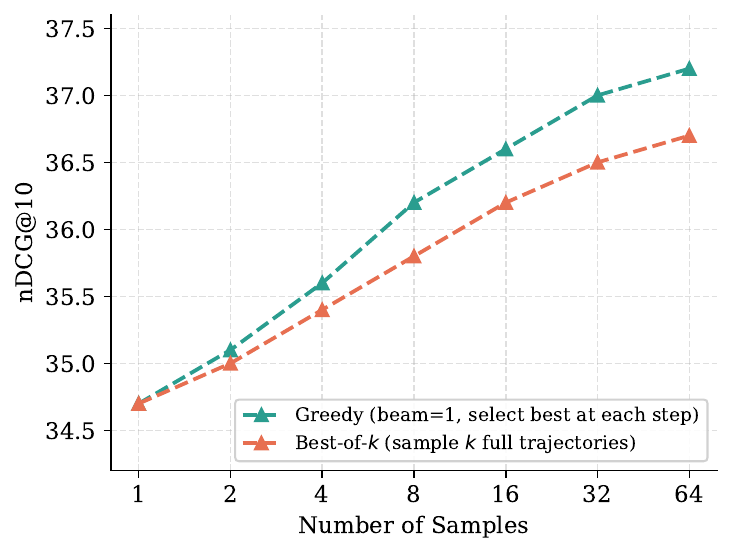}
\caption{Inference-time scaling of AURA on BRIGHT. Greedy selects the best continuation at each step using utility estimator; best-of-$k$ samples $k$ full trajectories and selects the one with highest estimated utility.}
\label{fig:inference_scaling}
\end{figure}

Figure~\ref{fig:inference_scaling} summarizes the performance-compute tradeoff on BRIGHT. Increasing search budget improves retrieval performance for both strategies, suggesting that additional reasoning can translate into stronger retrieval representations when guided by utility estimation. However, greedy expansion consistently achieves better compute efficiency than trajectory-level Best-of-$k$ search. Performance improves from 34.7 ($k=1$) to 37.2 ($k=64$), compared to 36.7 under Best-of-$k$ on the same search scale.
These results suggest that retrieval-oriented reasoning benefits more from early utility-guided decisions than from selecting among completed reasoning trajectories. By prioritizing promising intermediate states throughout the reasoning process rather than evaluating trajectories only after completion, AURA allocates inference-time computation more effectively.

\begin{table}[t]
\centering
\small
\caption{Inference-time reasoning compute on BRIGHT. }
\label{tab:compute_efficiency}
\begin{adjustbox}{width=0.99\linewidth}
\begin{tabular}{lcccc}
\toprule
\textbf{Method} & \textbf{Tokens} & \textbf{Steps} & \textbf{Latency (s)} & \textbf{nDCG@10} \\
\midrule
\textbf{AURA (utility estimator)} & \textbf{96} & \textbf{2.8} & \textbf{0.76} & \textbf{34.7} \\
AURA (full trajectory) & 332 & 9.5 & 4.60 & 33.0 \\
DIVER + GPT-4$^\dagger$  & 479 & 9.6 & --- & 32.1 \\
\bottomrule
\end{tabular}
\end{adjustbox}
\begin{flushleft}
\scriptsize $^\dagger$ Uses a separate GPT-4 model for query expansion; AURA generates reasoning within its own 4B backbone without external models.
\end{flushleft}
\end{table}

\subsection{Qualitative Analysis}
\label{sec:qualitative}
Trajectory inspection shows that discarded later steps can be semantically valid rather than incorrect. In the insects-and-LEDs example in \cref{fig:concept}, early reasoning identifies the proximate-versus-ultimate causation distinction, improving separation from hard negatives. Later steps add evolutionary and sensory-ecology background. Though useful, these additions repeat encoded constraints or generalize the query toward hard negatives about evolution and insect ecology. The estimator therefore stops once the discriminative constraint is encoded. This illustrates AURA's central distinction: informative text need not provide retrieval utility; stopping should depend on document discrimination rather than textual detail.

\section{Conclusion}
We presented AURA, a utility-guided framework that supervises intermediate reasoning states and stops generation when predicted retrieval utility ceases to improve. AURA achieves the strongest average BRIGHT performance among the compared methods while substantially reducing reasoning cost, without an external inference-time reasoner. Controlled ablations and full-corpus analysis show that the gains arise from utility-aware supervision rather than denser supervision or trajectory filtering, and that retrieval specificity closely tracks nDCG@10. Results on MS MARCO suggest limited in-domain degradation, while utility-guided greedy expansion further improves the performance--compute trade-off.

\section*{Limitations}
Despite its efficiency gains, AURA remains slower than single-pass retrieval because reasoning is generated autoregressively. Its utility labels are constructed once from a warmed-up checkpoint and may become less accurate as the embedding space evolves during subsequent optimization. Finally, our evaluation focuses primarily on reasoning-intensive English retrieval; broader validation on web search, multilingual retrieval, and generation tasks is needed to establish generality and characterize the remaining retrieval--generation trade-off.

\section*{Acknowledgments}
Yingtao Luo received support from the OpenAI Researcher Access Program. Shiyi Du was partially supported by a SoftBank Group/Arm Ph.D. Fellowship.
AI-assisted tools were used to improve language and presentation. The authors reviewed and take full responsibility for all technical claims and final content.

\bibliography{custom}

\clearpage

\end{document}